\documentclass[11pt,reqno]{amsart}
\usepackage{amssymb,epsf}
\usepackage{amstext,amsmath,amsfonts,amscd,amsthm,graphicx}
\theoremstyle{amsart}
\newtheorem{theorem}{Theorem}
\begin{document}

\title[]{\protect\vspace*{-30mm}On complete integrability of the Mikhailov--Novikov--Wang system}
\author[]{Petr Voj\v{c}\'{a}k}
\address{Mathematical Institute, Silesian University in Opava, Na Rybn\'{i}\v{c}ku 1, 746 01 Opava, Czech Republic}
\email{Petr.Vojcak@math.slu.cz}
\begin{abstract}
We obtain compatible Hamiltonian and symplectic structure for a new two-component fifth-order integrable system recently found by Mikhailov, Novikov and Wang, and show that this system possesses a hereditary recursion operator and infinitely many commuting symmetries and conservation laws, as well as infinitely many compatible Hamiltonian and symplectic structures, and is therefore completely integrable.
The system in question admits a reduction to the Kaup--Kupershmidt equation.\\

\end{abstract}

\maketitle

In the course of an ongoing classification of integrable polynomial evolution systems in two independent and two dependent variables Mikhailov, Novikov and Wang \cite{MacCallum} (see also \cite{Mikhailov}) have found a system
\begin{eqnarray}
\label{MNW_system}
\begin{array}{ccl}
u_t & = & -\frac{5}{3}u_5-10vv_3-15v_1v_2+10uu_3+25u_1u_2-6v^2v_1+6v^2u_1\\
& & +12uvv_1-12u^2u_1,\\
v_t & = & 15v_5+30v_1v_2-30v_3u-45v_2u_1-35v_1u_2-10vu_3-6v^2v_1\\
& & +6v^2u_1+12u^2v_1+12vuu_1.
\end{array}
\end{eqnarray}
Here and below $u_i=\frac{\partial^i u}{\partial x^i}$, $v_j=\frac{\partial^j v}{\partial x^j}$. Note that (\ref{MNW_system}) is one of just two new integrable systems found in \cite{MacCallum}, and therefore it is natural to explore its properties in order to find out whether it enjoys any features not present in the other higher-order integrable systems.

Upon setting $v\equiv 0$ the system (\ref{MNW_system}) reduces \cite{MacCallum} to the well-known Kaup--Kupershmidt equation, see e.g. \cite{Kaup, Wang} and references therein for more details on the latter.

Using the so-called symbolic method Mikhailov et al.\ \cite{MacCallum} proved that the system (\ref{MNW_system}) possesses infinitely many generalized symmetries (in the sense of \cite{Olver}) of orders $m\equiv 1,5 \mod 6$.
However, this result alone neither provides an explicit construction for the symmetries in question nor does it necessarily entail the existence of a recursion operator, see e.g.\ \cite{BSW, Sergyeyev2} and references therein. On the other hand, no
recursion operator, symplectic or (bi-)Hamiltonian structure for (\ref{MNW_system}) was found so far.
In view of the important role played by these quantities in establishing integrability, see e.g.\
\cite{ Blaszak, Olver} and references therein,
it is natural to ask whether (\ref{MNW_system}) admits any such quantities at all, and it is the goal of the present paper to show that this is indeed the case and thus (\ref{MNW_system}) indeed is a completely integrable system.

To this end we have first found a few low-order symmetries and cosymmetries of (\ref{MNW_system}) and subsequently constructed nonlocal parts of the operators in question (and then the operators {\it per se}) using these quantities, cf.\ e.g.\ \cite{Kersten, Marvan, Sanders}. As a result, we arrive at the following assertion.\looseness=-1

\begin{theorem}\label{thm}
The system (\ref{MNW_system}) possesses a Hamiltonian operator
\begin{eqnarray}
\label{hamiltonian}
\mathcal{P}=\begin{pmatrix} D_x^3-\frac{6}{5}uD_x-\frac{3}{5}u_1 & -\frac{6}{5}vD_x-\frac{3}{5}v_1\\ -\frac{6}{5}vD_x-\frac{3}{5}v_1 & 3D_x^3-(\frac{18}{5}u+\frac{12}{5}v)D_x-\frac{9}{5}u_1-\frac{6}{5}v_1 \end{pmatrix},
\end{eqnarray}
a symplectic operator
\begin{eqnarray}
\label{symplectic}
\mathcal{S}=\begin{pmatrix}  S_{11}+\frac{6}{5}\gamma_{21}D_x^{-1}\circ \gamma_{11} +\frac{6}{5}D_x^{-1}\circ \gamma_{21} & S_{12}+\frac{6}{5}\gamma_{21}D_x^{-1}\circ \gamma_{12} +\frac{6}{5}D_x^{-1}\circ \gamma_{22}\\ S_{21}+\frac{6}{5}\gamma_{22}D_x^{-1}\circ \gamma_{11} & S_{22}+\frac{6}{5}\gamma_{22}D_x^{-1}\circ \gamma_{12}  \end{pmatrix},
\end{eqnarray}
and a hereditary recursion operator $\mathcal{R}=\mathcal{P} \circ \mathcal{S}$ that can be written as
\begin{eqnarray}
\label{recursion}
\mathcal{R}=\begin{pmatrix}  R_{11} +G_{11}D_x^{-1}\circ \gamma_{11} +G_{12}D_x^{-1}\circ \gamma_{21}  & R_{12} +G_{11}D_x^{-1}\circ\gamma_{12} +G_{12}D_x^{-1}\circ \gamma_{22}\\ R_{21} +G_{21}D_x^{-1}\circ \gamma_{11} +G_{22}D_x^{-1}\circ \gamma_{21} & R_{22} +G_{21}D_x^{-1}\circ \gamma_{12} +G_{22}D_x^{-1}\circ\gamma_{22} \end{pmatrix},
\end{eqnarray}
where\\

\noindent $S_{11}=-D_x^3+6uD_x+3u_1, \hspace{3mm} S_{12}=-6vD_x+3v_1,$\\

\noindent $S_{21}=-6vD_x-9v_1, \hspace{3mm} S_{22}=9D_x^3-(\frac{54}{5}u-\frac{36}{5}v)D_x-\frac{27}{5}u_1+\frac{18}{5}v_1,$\\

\noindent $\gamma_{11}=1, \hspace{3mm} \gamma_{12}=0, \hspace{3mm} \gamma_{21}=u_2-\frac{12}{5}u^2+\frac{6}{5}v^2, \hspace{3mm} \gamma_{22}=-\frac{6}{5}v^2+\frac{12}{5}uv-3v_2,$\\

\noindent $\begin{array}{ccl}\!\!\!R_{11}\!\!\!&=&\!\!\!D_x^6 -\frac{36}{5}uD_x^4-\frac{108}{5}u_1D_x^3-( \frac{147}{5}u_2-\frac{324}{25}u^2 +\frac{252}{25}v^2) D_x^2\\
 & &\!\!\!-(21u_3-\frac{216}{5}uu_1+36vv_1) D_x-\frac{39}{5}u_4+\frac{738}{25}uu_2-\frac{666}{25}vv_2+\frac{621}{25}u_1^2\\
 & &\!\!\!-\frac{423}{25}v_1^2-\frac{864}{125}u^3+\frac{864}{125}uv^2-\frac{216}{125}v^3,
\end{array}$\\

\noindent $\begin{array}{ccl}\!\!\!R_{12}\!\!\!&=&\!\!\!\frac{84}{5}vD_x^4+\frac{102}{5}v_1D_x^3+(\frac{63}{5}v_2-\frac{576}{25}uv+\frac{252}{25}v^2)D_x^2\\
& &\!\!\!+(\frac{21}{5}v_3+\frac{576}{25}vv_1-\frac{144}{5}vu_1-\frac{396}{25}uv_1)D_x-\frac{216}{125}uv^2+\frac{432}{125}u^2v+\frac{36}{5}vv_2\\
& &\!\!\!-\frac{234}{25}u_2v-\frac{18}{5}uv_2-\frac{36}{5}u_1v_1+\frac{126}{25}v_1^2+\frac{3}{5}v_4,
\end{array}$\\

\noindent $\begin{array}{ccl}\!\!\!R_{21}\!\!\!&=&\!\!\!\frac{84}{5}vD_x^4+\frac{402}{5}v_1D_x^3-(\frac{576}{25}uv-\frac{729}{5}v_2+\frac{252}{25}v^2)D_x^2\\
& &\!\!\!-(\frac{648}{25}vu_1+\frac{1908}{25}uv_1-\frac{657}{5}v_3+\frac{432}{25}vv_1)D_x+\frac{216}{125}v^2u-\frac{1782}{25}uv_2-\frac{108}{25}vv_2\\
& &\!\!\!-\frac{486}{25}vu_2+\frac{297}{5}v_4+\frac{378}{25}v_1^2+\frac{432}{125}u^2v-\frac{1656}{25}u_1v_1,
\end{array}$\\

\noindent $\begin{array}{ccl}\!\!\!R_{22}\!\!\!&=&\!\!\!-27D_x^6+\frac{324}{5}uD_x^4+(\frac{648}{5}u_1-\frac{324}{5}v_1)D_x^3+(\frac{252}{25}v^2-\frac{972}{25}u^2-\frac{486}{5}v_2+\frac{729}{5}u_2)D_x^2\\
& &\!\!\!+(81u_3-54v_3-\frac{1944}{25}uu_1+\frac{684}{25}vv_1-\frac{648}{25}vu_1+\frac{648}{25}uv_1)D_x-\frac{486}{25}uu_2+\frac{432}{125}uv^2\\
& &\!\!\!-\frac{324}{25}vu_2+\frac{324}{25}uv_2+\frac{198}{25}vv_2-\frac{54}{5}v_4+\frac{153}{25}v_1^2-\frac{243}{25}u_1^2+\frac{81}{5}u_4-\frac{216}{125}v^3,
\end{array}$\\

\noindent $G_{11}=-\frac{6}{5}u_5-\frac{36}{5}vv_3-\frac{54}{5}v_1v_2+\frac{36}{5}uu_3+18u_1u_2-\frac{108}{25}v^2v_1+\frac{108}{25}v^2u_1+\frac{216}{25}uvv_1-\frac{216}{25}u^2u_1,$\\

\noindent $\begin{array}{ccl}\!\!G_{21}\!\!\!&=&\!\!\!\frac{54}{5}v_5+\frac{108}{5}v_1v_2-\frac{108}{5}v_3u-\frac{162}{5}v_2u_1-\frac{126}{5}v_1u_2-\frac{36}{5}vu_3-\frac{108}{25}v^2v_1+\frac{108}{25}v^2u_1\\
& &\!\!\!+\frac{216}{25}u^2v_1+\frac{216}{25}vuu_1,
\end{array}$\\

\noindent $G_{12}=\frac{18}{25}u_1, \hspace{3mm} G_{22}=\frac{18}{25}v_1.$
\end{theorem}

This result can be readily verified by straightforward but tedious computation.

Note that we can write the nonlocal terms of recursion operator (\ref{recursion}) and symplectic operator (\ref{symplectic}) in the form (cf. \cite{Maltsev, Sanders, Sergyeyev})
\begin{eqnarray}
\mathcal{R}_{-}=\sum\limits_{\alpha=1}^{2} G_{\alpha} \otimes D_x^{-1} \circ \gamma_{\alpha},\\
\mathcal{S}_{-}=\frac{6}{5}(\gamma_{1}^T \otimes D_x^{-1} \circ \gamma_{2}+\gamma_{2}^T \otimes D_x^{-1} \circ \gamma_{1}),
\end{eqnarray}
where
$$G_{\alpha}=\left( \! \begin{array}{c}
G_{1\alpha}\\
G_{2\alpha}
\end{array} \! \right), \hspace{5mm}
\gamma_{\alpha}=(\!\! \begin{array}{cc}
\gamma_{\alpha 1},\!\!\! & \gamma_{\alpha 2}
\end{array}\!\! ).$$
This enables one to rewrite the quantities $\mathcal{P}$, $\mathcal{S}$ and $\mathcal{R}$ in the so-called Guthrie form \cite{Guthrie, Artur}.

Let us also mention that the recursion operator $\mathcal{R}$ can be employed for the construction of a zero-curvature representation for (\ref{MNW_system}), and we intend to address this issue in our future work.

Setting $v \equiv 0$ in $\mathcal{P}$, $\mathcal{S}$, and $\mathcal{R}$ and extracting the appropriate entries thereof we readily recover the Hamiltonian, symplectic and recursion operator for the Kaup--Kupershmidt equation, cf.\ e.g.\ \cite{Wang} and references therein.

Moreover, it readily follows from Theorem~\ref{thm} that the system (\ref{MNW_system}) has, as usually is the case for integrable systems (see e.g. \cite{Blaszak}), infinite hierarchies of compatible Hamiltonian operators $\mathcal{R}^k \circ \mathcal{P}$ and symplectic operators $\mathcal{S} \circ \mathcal{R}^k$, $k=0,1,2 \ldots$ In particular, this means that (\ref{MNW_system}) is a multi-Hamiltonian system.

While the Hamiltonian operator $\mathcal{P}$ is local, it is straightforward to verify that  all Hamiltonian operators of the form $\mathcal{R}^k \circ \mathcal{P}$, $k=1,2,\dots$, are nonlocal. We conjecture that $\mathcal{P}$ is the only local Hamiltonian structure for the Mikhailov--Novikov--Wang system (1). Note also that all symplectic structures $\mathcal{S} \circ \mathcal{R}^k$, $k=0,1,2,\dots$, including $\mathcal{S}$ itself, are nonlocal.

Furthermore, it is possible to construct two infinite sequences of conserved functionals $H_{1,k}$
and $H_{2,k}$ given by the formula
\begin{equation}\label{Hamiltonians}
\delta H_{i,k}=(\mathcal{R}^{\ast})^k (\delta H_i),
\end{equation}
where
$$H_1=-\frac{5}{3} \int u dx,\qquad  H_2= \int \left( \frac{5}{6}u_1^2-\frac{5}{2}v_1^2+\frac{4}{3}u^3+\frac{2}{3}v^3-2uv^2 \right) dx,$$
$\mathcal{R}^{\ast}=\mathcal{S}\circ\mathcal{P}$ is the formal adjoint of $\mathcal{R}$ and $\delta$ stands for the variational derivative. It is readily seen that all functionals $H_{i,k}$ are in involution with respect to Poisson brackets associated with the Hamiltonian structures $\mathcal{R}^s \circ \mathcal{P}$ for all $s=0,1,2,\dots$.
By Proposition 2 of \cite{Sergyeyev}, for all functionals $H_{i,k}\equiv\int\rho_{i,k}dx$, $i=0,1,2,\dots$, $k=1,2$, 
their densities $\rho_{i,k}$ defined recursively through (\ref{Hamiltonians}) are local, i.e.\ they depend (at most) on $x,t,u,v$ and a finite number of $u_j$ and $v_j$. 
System (\ref{MNW_system}) can be written in the Hamiltonian form
\begin{equation}\label{Ham}\left( \! \begin{array}{c}
u_t\\
v_t
\end{array} \! \right)=\mathcal{P}\delta H_2.
\end{equation}

Let
\begin{equation}\label{Ham2}
Q_1=\left( \! \begin{array}{c}
u_1\\
v_1
\end{array} \! \right)=\mathcal{P}\delta H_1, \quad Q_2=\mathcal{P}\delta H_2,
\end{equation}
i.e. $Q_2$ is a column containing the right-hand sides of (\ref{MNW_system}). The recursion operator (\ref{recursion}) and the symmetries with the characteristics $Q_1$ and $Q_2$ are readily verified to meet the requirements of Theorem 1 from \cite{Sergyeyev}, and therefore the symmetries with the characteristics $Q_{i,j}=\mathcal{R}^j(Q_i)$, $i=1,2$, $j=0,1,2,\ldots$, are local. 
In fact, it can be shown that for any given $i$ and $j$ the characteristic $Q_{i,j}$ depends only on $u,v,u_1,v_1,\ldots,u_{1+4(i-1)+6j},v_{1+4(i-1)+6j}$. Moreover, as the recursion operator (\ref{recursion}) is hereditary and the symmetries with the characteristics $Q_1$ and $Q_2$ commute, so do the symmetries with characteristics $Q_{i,j}$ for all $i=1,2$ and all $j=0,1,2,\ldots$

For example, seventh-order symmetry of (\ref{MNW_system}) has the characteristic $Q_{1,1}=(Q_{1,1}^1,Q_{1,1}^2)^T$, where
\begin{eqnarray}
\label{7_order}
\begin{array}{ccl}
Q_{1,1}^1\!\!\!&=&\!\!\! u_7-\frac{42}{5}uu_5+\frac{126}{5}u_1^3-\frac{126}{5}u_1v_1^2-\frac{147}{5}u_1u_4-\frac{1386}{25}v_2u_1v+\frac{2268}{25}uu_1u_2\\
& & \!\!\!+\frac{1512}{125}uv^2u_1-\frac{2016}{125}u^3u_1-\frac{252}{125}v^3u_1+\frac{1512}{125}v_1vu^2+\frac{504}{25}u^2u_3-\frac{756}{25}uvv_3\\
& & \!\!\!-\frac{756}{125}v^2uv_1-\frac{756}{25}uv_1v_2+\frac{252}{25}v^2v_3+\frac{84}{5}v_2v_3+21v_1v_4+\frac{84}{5}vv_5-\frac{252}{5}u_2u_3\\
& & \!\!\!-\frac{252}{25}v^2u_3+\frac{126}{25}v_1^3-\frac{1134}{25}v_1u_2v+\frac{756}{25}vv_1v_2,\\
Q_{1,1}^2\!\!\!&=&\!\!\! -27v_7-\frac{756}{25}v_2u_1v+\frac{252}{25}v^2v_3+\frac{126}{25}v_1^3+189v_4u_1+\frac{483}{5}u_4v_1-\frac{378}{5}u_1^2v_1\\
& & \!\!\!+\frac{1386}{5}u_2v_3-\frac{756}{5}v_2v_3-\frac{252}{125}v^3v_1-\frac{756}{25}uvu_3-\frac{756}{25}v_1u_2v+\frac{1512}{25}uv_1v_2\\
& & \!\!\!+\frac{882}{25}vv_1v_2-\frac{1134}{25}u_1vu_2+\frac{378}{25}u_1v_1^2+\frac{378}{5}v_5u-\frac{252}{25}v^2u_3-\frac{3024}{25}u_2uv_1\\
& & \!\!\!-\frac{378}{5}v_1v_4-\frac{4536}{25}v_2u_1u+\frac{1008}{125}u^3v_1+\frac{756}{125}uv^2u_1-\frac{1512}{25}u^2v_3+\frac{1134}{5}u_3v_2\\
& & \!\!\!+\frac{1512}{125}u^2u_1v+\frac{84}{5}u_5v.
\end{array}
\end{eqnarray}

Let us also mention that the formulas (\ref{Ham}) and (\ref{Ham2}) are special cases of a more general relation
\begin{equation}\label{Ham3}
Q_{i,j}=\mathcal{P}\delta H_{i,j},\quad i=1,2,\quad j=0,1,2,\dots,
\end{equation}
and the flows associated with $Q_{i,j}$ with $j\geqslant 1$ are bi-Hamiltonian, as we have
\begin{equation}\label{Ham4}
Q_{i,j}=\mathcal{P}\delta H_{i,j}=\tilde{\mathcal{P}}\delta H_{i,j-1},\quad i=1,2,\quad j=1,2,\dots.
\end{equation}
by virtue of (\ref{Hamiltonians}). Here $\tilde{\mathcal{P}}=\mathcal{R}\circ\mathcal{P}$.


\section*{Acknowledgments}

The author thanks Dr.\ A. Sergyeyev for useful discussions. This research was supported by Silesian University in Opava under the student grant project SGS/18/2010 and by the Ministry of Education, Youth and Sports of Czech Republic under the grant MSM4781305904.\\

\end{document}